\documentclass[journal, a4paper]{IEEEtran}

\IEEEoverridecommandlockouts

\usepackage{units}
\usepackage{graphicx}
\usepackage{multirow}
\usepackage{amsmath,amssymb,latexsym}
\usepackage{rotating}
\usepackage{lettrine}
\usepackage{hyperref}

\newcommand{\Ntx}{\mathrm{N}_\mathrm{TX}}
\newcommand{\Nrx}{\mathrm{N}_\mathrm{RX}}
\newcommand{\Ntries}{N_\mathrm{tries}}
\newcommand{\cell}[1]{\Ntx\!\xrightarrow[]{\!#1\!}\!\Nrx}
\newcommand{\cellr}[1]{\Nrx\!\xrightarrow[]{\!#1\!}\!\Ntx}
\newcommand{\frm}[1]{\mathcal{F}_{#1}}
\newcommand{\frms}[1]{\mathcal{F}^{.s}_{#1}}
\newcommand{\frmsx}[2]{\mathcal{F}^{.s=#2}_{#1}}
\newcommand{\ack}[1]{\mathcal{A}_{#1}}
\newcommand{\epsf}{\epsilon_\mathrm{f}}
\newcommand{\epsa}{\epsilon_\mathrm{a}}
\newcommand{\Tapp}{T_\mathrm{app}}
\usepackage{xcolor}
\usepackage{alltt}

\usepackage{textcomp}

\usepackage[english]{babel}
\usepackage{lipsum}% For 'Lorem Ipsum' dummy text
\usepackage[pscoord]{eso-pic}% The zero point of the coordinate systemis the lower left corner of the page (the default).

\newcommand{\placetextbox}[3]{% \placetextbox{<horizontal pos>}{<vertical pos>}{<stuff>}
  \setbox0=\hbox{#3}% Put <stuff> in a box
  \AddToShipoutPictureFG*{% Add <stuff> to current page foreground
    \put(\LenToUnit{#1\paperwidth},\LenToUnit{#2\paperheight}){\vtop{{\null}\makebox[0pt][c]{#3}}}%
  }%
}%

\title{Enhanced Energy-Saving Mechanisms in TSCH Networks for the IIoT: the PRIL Approach
\thanks{Stefano Scanzio, Gianluca Cena, and Adriano Valenzano are with the National Research Council of Italy (CNR--IEIIT), Corso Duca degli Abruzzi 24, I-10129 Torino, Italy (Corresponding author: Stefano Scanzio)}
}

\author{Stefano Scanzio, \textit{Senior Member}, \textit{IEEE}, Gianluca Cena, \textit{Senior Member}, \textit{IEEE},\\ and Adriano Valenzano, \textit{Senior Member}, \textit{IEEE}
}

\begin{document}

\placetextbox{0.5}{1}{This article has been accepted for publication in IEEE Transactions on Industrial Informatics.}
\placetextbox{0.5}{0.985}{This is the author's version which has not been fully edited and
  content may change prior to final publication.}
\placetextbox{0.5}{0.97}{Citation information: DOI \href{https://dx.doi.org/10.1109/TII.2022.3209258}{10.1109/TII.2022.3209258}}%
\placetextbox{0.5}{0.05}{This work is licensed under a Creative Commons Attribution-NonCommercial-NoDerivatives 4.0 License.}
\placetextbox{0.5}{0.035}{For more information, see https://creativecommons.org/licenses/by-nc-nd/4.0/}%

\maketitle
\thispagestyle{empty}
\pagestyle{empty}

\begin{abstract}
 Lifetime of motes in wireless sensor networks can be enlarged by decreasing the energy spent for communication. Approaches like time slotted channel hopping pursue this goal by performing frame exchanges according to a predefined schedule, which helps reducing the duty cycle. Unfortunately, whenever the receiving radio interface is active but nobody in the network is transmitting, idle listening occurs. If the traffic pattern is known in advance, as in the relevant case of periodic sensing, proactive reduction of idle listening (PRIL) noticeably lowers energy waste by disabling receivers when no frames are expected for them. Optimal PRIL operation demands that, at any time, the transmitter and receiver sides of a link have a coherent view of its state (either enabled or disabled). However, this is not ensured in the presence of acknowledgment frame losses.

 This paper presents and analyzes some strategies to cope with such events. An extensive experimental campaign has been carried out through discrete event simulation to determine what consequences above errors may have from both a functional and performance viewpoint. Results show that, although no strategy is optimal in all circumstances, different solutions can be profitably adopted depending on the specific operating conditions.
\end{abstract}


%%%%%%%%%%%%%%%%%%%%%%%%%%%%%%%%%%%%%%%%%%%%%%%%%%%%%%%%%%%%%%%%%%%%%%%%%%%%%%%%

\section{Introduction}
\label{sec:introduction}
\lettrine[lines=2, findent=0.2em, nindent=0em]{I}n the past decade 
wireless extensions were progressively integrated into the wired infrastructure of industrial distributed systems.
This results in current heterogeneous networks, which rely on different transmission technologies \cite{8819994}.
In particular, wireless sensor (and actuator) networks (WSN/WSAN) are increasingly employed wherever the lack of a suitable communication infrastructure makes wireless meshing convenient.
Besides the ability to satisfy the timing constraints imposed by applications (e.g., deadlines), 
ultra-low power consumption is often demanded.
The ability not to waste energy 
is perhaps the primary
requirement of wireless networks whose nodes are powered on batteries and must operate unattended for long time periods, often exceeding ten years. 

WSN/WSAN are core technologies for the Industrial Internet of Things (IIoT) \cite{9381665,8401919}.  
Significant examples of how they can be exploited in such context are the 
retrofitting of industrial plants for monitoring and diagnostics \cite{2019-IA-fault_diagnosis}, 
greenhouse automation \cite{kolapkar2020greenhouse}, 
heating, ventilation, and air conditioning (HVAC) systems \cite{9548990}, 
smart and precision agriculture \cite{9122412}, petrochemical industry \cite{zhao2021research} and many others.
Several commercial solutions are based on the well-know IEEE 802.15.4 standard \cite{IEEE-802.15.4-2020}, 
including its Time Slotted Channel Hopping (TSCH) operating mode. 
TSCH relies on scheduled transmissions and mechanisms to make the network nodes sleep when they are not involved in the communication, 
thus allowing tangible energy saving \cite{9187609}.
As a side effect, this also reduces the amount of exhausted batteries to be disposed of and improves sustainability.
Popular examples of TSCH-based technologies that are customarily used in industrial environments are ANSI/ISA 100.11a and IEC WirelessHART \cite{PETERSEN2011}, as well as IETF 6TiSCH \cite{8823863}.

In TSCH, scheduled transmissions between a pair of nodes 
(that is, opportunities for the sender node to use a reserved portion of the available bandwidth) 
repeat cyclically over time, with a period that is typically $\unit[2]{s}$ or less. 
Receiver nodes are activated accordingly to read incoming frames. 
When a receiver is switched on but no transmissions take place, a condition commonly known as \textit{idle listening} occurs,
which causes non-negligible energy waste.
However, if the traffic pattern generated by applications can be predicted to some extent, 
as in the case of periodic sampling performed at the IIoT perception layer,
techniques known as Proactive Reduction of Idle Listening (PRIL) permit to limit this phenomenon dramatically,
by temporarily disabling the receiver on a link when no frames are expected to be exchanged soon.

A preliminary PRIL scheme, called PRIL-F, was presented in \cite{PRIL-F},
where the transmitter puts the receiver to sleep by including suitable commands along with data.
Such mechanism proved to offer significant energy saving in specific network topologies often found in industrial control systems, 
for instance the single-hop star topology. 
In that work, the possibility that an acknowledgement (ACK) frame could be lost was not taken into account. 
However, such events are not unlikely to happen, and may cause the two sides of the link to have an inconsistent view of its state.
For example, the receiver could be sleeping, but the transmitter is not aware of this and keeps on sending frames uselessly.
Since the effect of sleep commands is temporary, the correct TSCH operations are not impaired.
Nevertheless, negative consequences on energy saving capability are definitely expected.

This paper explicitly tackles ACK frame losses
by proposing and comparing some 
strategies to deal with their occurrence.
Besides, it also offers guidelines to select the most appropriate solution depending on the network operating conditions.
Compared to \cite{PRIL_etfa2021}, this work analyzes and properly solves the above problem 
by: 
\begin{enumerate}
\item deriving the actual failure probability for ACK frames from experiments performed with real devices;
\item proposing a new, quite effective \textit{A-open} strategy; and
\item analyzing alternative strategies, like \textit{2-open}, to show their lack of effectiveness.
\end{enumerate}
The structure of the paper is the following: 
Section~\ref{sec:PRIL} briefly reviews the literature about energy-saving techniques 
and summarizes some main characteristics of PRIL, 
Section~\ref{sec:ACK_loss} deals with the strategies mentioned above, 
while Sections~\ref{sec:experimental_setup} and \ref{sec:results} present the setup adopted in the experiments and the obtained results, respectively. 
Finally, Section~\ref{sec:conclusions} draws some concluding remarks.

\section{Energy Saving in TSCH}
\label{sec:PRIL}
Conserving energy is likely the main goal of WSN since their inception, and surely one of the main reasons behind TSCH.
Many works appeared in the literature focus on energy consumption in TSCH. 
Some of them adopt clever scheduling algorithms to assign time slots and frequencies, 
so as to extend the global network lifetime \cite{7962719, 8371162, 8931363}. 
When sporadic traffic is considered, 
other solutions apply specific scheduling algorithms to shared cells \cite{8115802}, 
that is, cells which can be used by more than one sender node.
An improved node load balancing is used in \cite{doi:10.1504/IJAHUC.2020.107505} to prolong network lifetime. 
Instead, \cite{MOHAMADI20221} aims at reducing energy consumption during network formation.
Works in \cite{s18103556,9187609} suggest practical guidelines and analyze how some communication parameter settings impact on  performance indicators, including power consumption.
The accuracy of synchronization among TSCH nodes is another aspect that affects performance and power consumption. This is why some works propose strategies to reduce \cite{s20216047} or dynamically adapt \cite{10.1007/978-3-030-30523-9_2} 
the guard time (a TSCH feature intended to cope with non-perfect synchronization), hence saving energy. 
Authors of \cite{s19194128} have shown how a dynamic modification of the enhanced beacon periods can help in improving synchronization and connection time, with a reduction of energy consumption.
Transmission channels can be selected on the fly by means of black and white listing techniques.  A better quality of the communication support at the medium access control (MAC) layer results in improvements to energy consumption, reliability, and latency at the same time \cite{label,9530419}.

PRIL techniques, on which this work relies, 
are typically orthogonal to those described above, 
and they are another appealing way to pursue the energy goal \cite{PRIL-F, PRIL_etfa2021}.
As mentioned above, cells reserved to the link between a sender ($\Ntx$) and a receiver ($\Nrx$) represent transmission opportunities.
Scheduled cells that are not actually used by $\Ntx$ for frame exchanges cause $\Nrx$ to listen uselessly. 
PRIL operates by switching $\Nrx$ off when the scheduled cells are expected to remain unused, 
based on predictions about the $\Ntx$ traffic.

TSCH nodes are time-synchronized, and time is divided into slots of \textit{fixed} duration $T_\mathrm{slot}$. 
In every slot several concurrent transmissions 
(up to the number $N_\mathrm{ch}$ of available channels) can be 
contextually performed by different senders. 
Usually, $N_\mathrm{ch}=16$ for the $\unit[2.4]{GHz}$ O-QPSK PHY defined by the IEEE 802.15.4 specification. 
Mapping between logical channels (channel offsets in the scheduling table) 
and physical channels (frequencies actually used for transmission) relies on a pseudo-random function. 
This technique, known as \textit{channel hopping}, improves robustness against disturbance thanks to frequency diversity, and results in higher reliability.

Every time slot is identified by the Absolute Slot Number (ASN), 
a counter shared among nodes that is incremented by one
on every slot. 
The protocol defines both \textit{shared} cells, which can be used by several nodes to transmit data,
and \textit{dedicated} (non-shared) cells, which can be only exploited by a specific sender. 
This paper explicitly focuses on confirmed (unicast) transmissions in dedicated cells,
which are by far the most interesting option for data exchanges (e.g., values cyclically sampled by sensors).
Although PRIL could be extended to deal with multicast transmissions and shared cells as well,
by permitting the transmitter to selectively switch off its associated receivers, doing so makes the mechanism more complex.
For this reason these cases have not been considered here, and are left as future work.

\begin{figure}[t]
	\begin{center}
	\includegraphics[width=1\columnwidth]{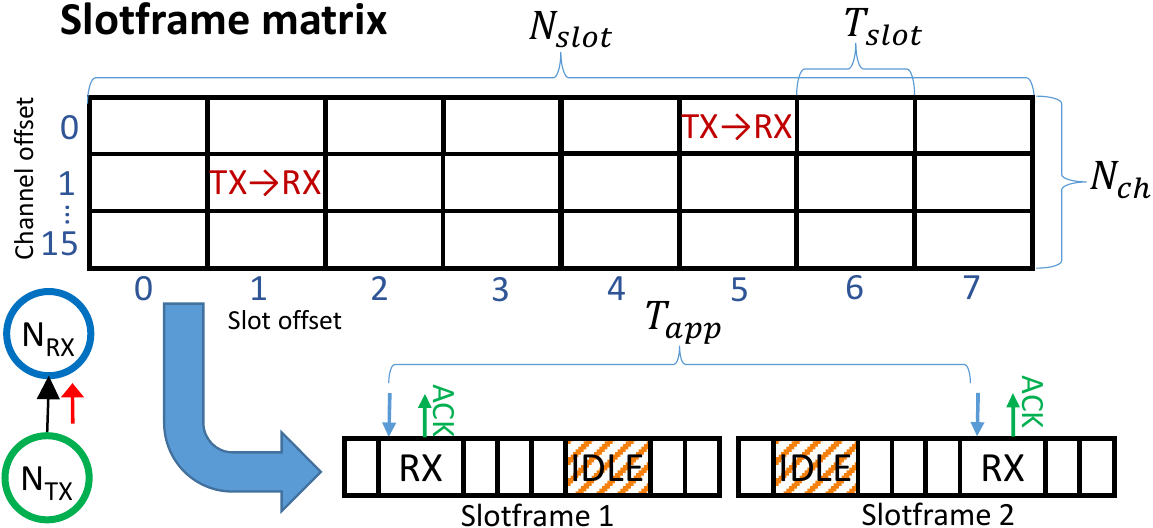}
	\end{center}
	\caption{Example of TSCH operation and compact view of the matrix.}
	\label{fig:TSCH}
\end{figure}

As outlined in the example in the upper part 
of Fig.~\ref{fig:TSCH}, a cell is defined in an $N_\mathrm{ch} \times N_\mathrm{slot}$ slotframe matrix
that repeats over time every $N_\mathrm{slot}$ slots. 
Thus, every scheduled cell provides node $\Ntx$ with a cyclic transmission opportunity that is made available with period 
$T_\mathrm{sfr} = N_\mathrm{slot} \cdot T_\mathrm{slot}$. The quantity $T_\mathrm{sfr}$ is typically known as slotframe interval.
Within the same matrix, more than one such opportunities can be defined for the same pair of nodes. 
This is the case of the \textit{link} from $\Ntx$ to $\Nrx$ in the figure, which has two.
The relative position of the current cell within the matrix (i.e., the column, which corresponds to the slot offset) 
is obtained by each node as $\mathrm{ASN}\ \mathrm{mod}\ N_\mathrm{slot}$. 
Channel hopping does not directly impact on the definition and implementation of PRIL techniques, 
and so a compact representation has been used in the following for the matrix, 
as shown at the bottom of Fig.~\ref{fig:TSCH}. 
The state of each scheduled cell is reported from the point of view of $\Nrx$, 
which is the only node directly affected by PRIL: 
either a frame is received in the cell (RX) 
or idle listening (IDLE) is experienced otherwise.
The use of PRIL has no effect on the quality of clock synchronization among nodes, 
because its implementation does not require any modification to the TSCH synchronization protocol.

Let $\overline T_\mathrm{app}$ be the mean time elapsing between two consecutive instances 
of a specific scheduled cell where a data frame is actually sent. 
In many industrial applications packet generation is periodic, in which case this time coincides with the sampling period $T_\mathrm{app}$. 
In typical configurations $\overline T_\mathrm{app} \gg T_{\mathrm{sfr}}$, and
the energy wasted for idle listening can be significant because, 
not considering retries, only one cell every 
$c \cdot {\overline T_\mathrm{app}}/{T_{\mathrm{sfr}}}$ is actually exploited for transmission 
($c$ is the link capacity, given by the number of cells in the matrix allocated to it).

Practical PRIL implementations rely on \textit{sleep commands} embedded by $\Ntx$ inside IEEE 802.15.4 Information Elements (IE). 
On the reception of a sleep command the $\Nrx$ receiving circuitry is switched off for a time period encoded in the command itself.
When the command effect ceases, the receiver automatically reverts to the conventional TSCH behavior.
Computing the duration of the sleep period for nodes located one hop away from the source node 
is a trivial task when traffic is periodic
(it is equal or directly related to $\Tapp$). 
PRIL-F \cite{PRIL-F} is based on this assumption and, to the best of our knowledge, 
it is the only PRIL implementation experimentally evaluated so far.
Other methods to evaluate the sleeping time in more complex scenarios, 
e.g., by leveraging traffic inspection and/or machine learning to predict cell usage patterns, are still under development.

In this work we consider cyclic data transmissions between two nodes with period $\Tapp$, which is coherent with PRIL-F. 
We wish to derive generic strategies that apply to 
any PRIL technique,
regardless of the way the sleeping time to be inserted in the sleep command is calculated.
Assuming the traffic to be periodic is not a big limitation, 
as most industrial applications that rely on WSN (and even home automation) work this way. 
Instead, restricting the analysis to a single hop has to do specifically with PRIL-F:
although it can be used also in multi-hop networks, 
power saving is only achieved on nodes located one hop away from the traffic source.
This approach has been proved able to lower power consumption sensibly, 
without affecting communication latency and reliability \cite{PRIL-F}, 
and consequently it can be profitably adopted in any context. 
PRIL strategies capable to reduce power consumption on nodes located more than one hop away from the traffic source 
are outside the scope of this paper and are left for future work. 
It is worth remarking that dynamic changes brought by RPL to the network topology
do not affect the correct operation of the network permanently, 
and not even of PRIL
(when moved to a new position, sleeping links are awakened automatically).

\begin{figure}[t]
	\begin{center}
	\includegraphics[width=0.9\columnwidth]{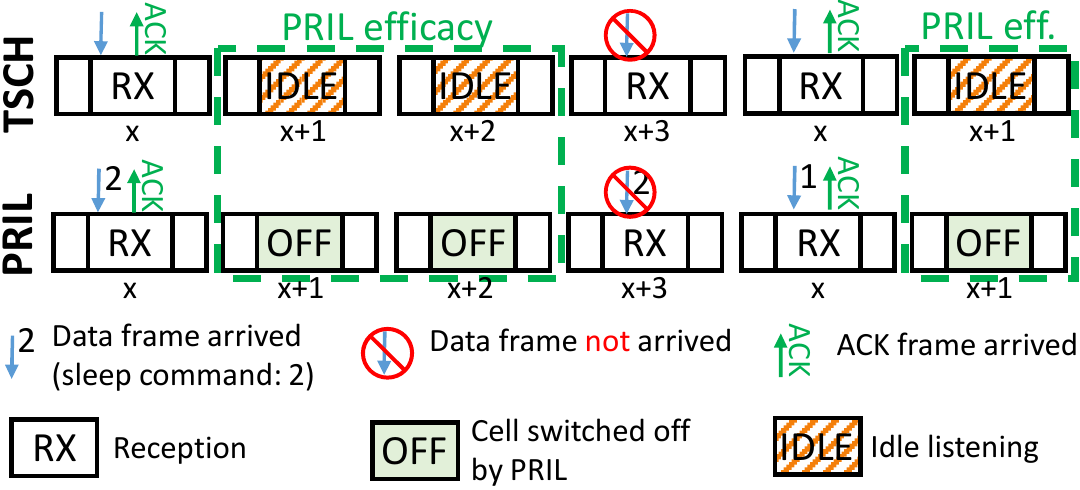}
	\end{center}
	\caption{Effectiveness of PRIL vs. conventional TSCH ($\Nrx$ viewpoint).}
	\label{fig:PRIL}
\end{figure}

Fig.~\ref{fig:PRIL} shows a simple example of PRIL operation. 
When a frame reaches $\Nrx$ (downward blue arrow in the figure), 
the receiver is
deactivated for a suitable time to prevent idle listening. 
In practice, the content of the sleep command is the number of scheduled cells that have to be ``disabled'' after command reception. 
When a data frame gets lost, for instance because of electromagnetic noise, 
some cells are used for re-transmissions,
and the sleeping time encoded in every retry is shortened accordingly.
The loss of an ACK frame is not considered in the example in Fig~\ref{fig:PRIL}.
Unfortunately, the analysis performed in Subsection~\ref{subsec:ACK_prob} on data acquired in real networks  shows that such a kind of event is not so rare, and
the probability to lose ACK frames is not much smaller than that for data frames. 
In the next section some techniques are presented that are explicitly aimed to deal with ACK losses.

\section{Robustness against ACK Losses}
\label{sec:ACK_loss}

\begin{figure*}[t]
	\begin{center}
	\includegraphics[width=2\columnwidth]{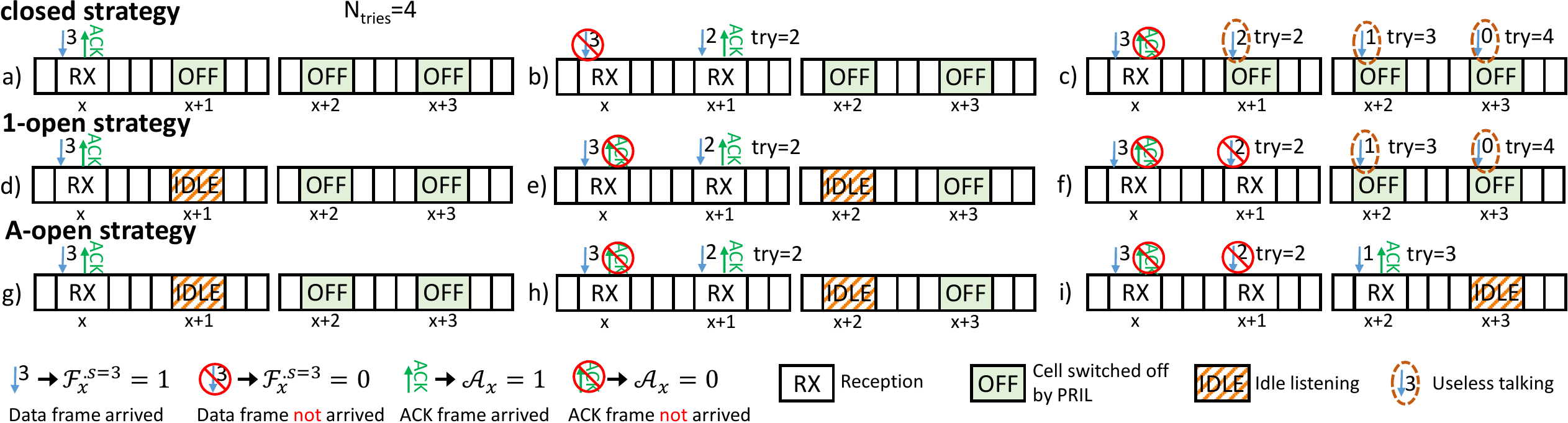}
	\end{center}
	\caption{Comparison among the behaviors of the considered PRIL strategies: \textit{closed}, \textit{1-open}, and \textit{A-open} ($N_{\mathrm{tries}}=4$).}
	\label{fig:comparison}
\end{figure*}

Let $\cell{}$ denote a \textit{reserved} cell devoted to transmissions between nodes $\Ntx$ and $\Nrx$. 
Notation $\cell{x}$ refers to the $x$-th occurrence of that cell,
where $x$ is a sequence number that identifies the specific cell instance. 
Thus, $\cell{x+1}$ represents the following occurrence. 
In the case only one scheduled cell is allocated to the pair of nodes, the next occurrence is 
characterized by the same slot offset but is found in the next slotframe. As said before, not all the scheduled cells are used for transmission, and
the fraction of them which are actually exploited depends on the value of $\Tapp$.
A reserved cell where the receiving interface of $\Nrx$ is switched off 
by PRIL is denoted $\cell{\overline{x}}$.

When a transmission takes place in $\cell{x}$, $\Nrx$ notifies $\Ntx$ of the correct reception of the (data) frame by means of the corresponding acknowledgement. 
In our analysis we associate events $\frm{x}$ and $\ack{x}$ to the transmission attempt for the frame and its acknowledgement, respectively.
With a little abuse of notation, 
we will use the same symbols $\frm{x}$ and $\ack{x}$ to denote 
both the specific data/acknowledgement frame
and the associated transmission outcome, when no ambiguity occurs.
We adopt the following convention: when the frame is successfully received by $\Nrx$ then $\frm{x}=1$. 
In this case, the related acknowledgment is either returned correctly ($\ack{x}=1$) or lost ($\ack{x}=0$). 
Instead, $\frm{x}=0$ means that the frame has not been received at all by $\Nrx$. 

It follows that $\Ntx$ is sure that the frame has been correctly received only when $\ack{x}=1$.
By contrast, event combinations $\frm{x}=0$ and $\frm{x}=1 \wedge \ack{x}=0$ are equivalent from the $\Ntx$ viewpoint,
despite they represent very different conditions for $\Nrx$. 
In particular, in both cases $\Ntx$ has to retransmit the same frame in the next cell ($\frm{x+1}$),
irrespective of whether it has already been delivered correctly to its destination.

The sleep command is modeled as an attribute $.s$ of the data frame ($\frms{x}$). 
For example, $\frmsx{x}{4}$ means that the frame embeds a sleep command whose duration is equal to $4$. 
It is intended to disable the next $4$ cells on the link 
($\cell{\overline{x+1,...,x+4}}$), 
so that transmission between $\Ntx$ and $\Nrx$ is re-enabled in $\cell{x+5}$. 
Using relative times in sleep commands simplifies the description of our techniques, 
however all proposed strategies can be easily implemented by using absolute times.

\subsection{Closed strategy}
The \textit{closed} strategy corresponds to the PRIL operation described in \cite{PRIL-F}.
Examples of its behavior are sketched in the upper part of Fig.~\ref{fig:comparison}.
In particular, in Fig.~\ref{fig:comparison}.c 
a frame that conveys a sleep command for deactivating the next $t$ cells
is assumed to be delivered correctly in cell $x$ ($\frmsx{x}{t}=1$), which causes the receiver to switch off cells $\cell{\overline{x+1,...,x+t}}$. 
However, that transmission attempt suffers from an ACK loss ($\ack{x}=0$).
Sender $\Ntx$, which misses the confirmation, retransmits the data frame, and keeps doing so
since the $\Nrx$ receiving interface is off. 
This condition, where the sender repeatedly fails to deliver frames because communication is disabled on the receiver,
is denoted \textit{useless talking}.

Let $\Ntries{}$ be the maximum number of transmission attempts the MAC is allowed to perform for a frame (retry limit plus one).
When $t \geq \Ntries{}-1$ all retries will unavoidably fail,
otherwise the link $\cell{}$ is re-enabled on $\Nrx$ 
after staying ``closed'' for $t$ reserved cells.
In both cases the loss of $\ack{x}$ results in an energy waste on $\Ntx$ 
caused by its useless talking that is proportional to $\min(\Ntries{}-1, t)$.
On the receiver side the energy saved by PRIL is optimal,
since the $\Nrx$ interface is switched off as soon as $\frmsx{x}{t}$ is received.
Every failed attempt possibly occurring at the beginning of any frame transmission 
($\frmsx{x+i}{t-i}=0 \mid_{i=0,...,\varphi-1}$, 
with $\varphi \geq 1$), 
causes both $t$ and the available retries to be decreased by one,
as shown in Fig.~\ref{fig:comparison}.b.
Such early failures reduce the amount of useless talking,
should they be followed by an ACK loss, 
because the energy spent in the related $\varphi$ cells is ascribed to transmission errors.

\subsection{1-open and n-open strategies}
The \textit{1-open} strategy lessens the limits of the \textit{closed} strategy by letting ``open'' the next reserved cell following the sleep command $\frmsx{x}{t}$ (Fig~\ref{fig:comparison}.d). 
In other words, after a correctly received frame ($\frmsx{x}{t}=1$), 
$\Nrx$ sets cell $\cell{x+1}$ as usable, thus deferring the shut-off of the receiving interface to cells $\cell{\overline{x+2,...,x+t}}$. 
By doing so, if $\ack{x}=0$ and the retry limit has not been exceeded
there is a chance for the next frame exchange to succeed ($\frmsx{x+1}{t-1}=1 \wedge \ack{x+1}=1$),
so that $\Ntx$ can stop unnecessary retransmissions (Fig~\ref{fig:comparison}.e). 
In this case, when $\frmsx{x+1}{t-1}=1$ the following reserved cell $\cell{x+2}$ is usable
and can be actually exploited if either $\ack{x+1}=0$ or other frames addressed to $\Nrx$ are queued in $\Ntx$.
The only drawback occurs when $\ack{x}=1$ and the $\Ntx$ queue is empty, because cell $\cell{x+1}$ remains unused 
and $\Nrx$ wastes energy in idle listening.

As in the \textit{closed} strategy, 
when an ACK loss is followed by a frame loss in the next reserved cell 
($\ack{x}=0 \wedge \frmsx{x+1}{t-1}=0$), 
$\Ntx$ keeps on transmitting and wastes energy in useless talking (Fig~\ref{fig:comparison}.f),
since $\Nrx$ is unable to send confirmations and stop automatic retransmission.
To mitigate this phenomenon,
\textit{n-open} strategies can be devised where, 
following a successful frame delivery, $n$ reserved cells are left open on $\Nrx$ to deal with possible retransmissions. 
Such strategies prevent the occurrence of conditions like the one depicted in Fig~\ref{fig:comparison}.f, 
and are able to cope with the loss of up to $n-1$ data frames following an acknowledgement loss 
($\ack{x}=0 \wedge \frm{x+1}=\frm{x+2}=...=\frm{x+n-1}=0$).

The probability to exceed the \textit{n-open} recovery capabilities, which 
depends on the failure probability for $\ack{}$ and $\frm{}$
($\epsa$ and $\epsf$, respectively) is typically small and equal to 
$\epsa \cdot \epsf^n$. 
For example, when $\epsa=\unit[8.0]{\%}$ and $\epsf=\unit[12.6]{\%}$ this probability is $\unit[0.13]{\%}$ for $n=2$, 
and drops to $\unit[0.016]{\%}$ for $n=3$. 
The cost for preventing these unlikely events is to accept idle listening in $n$ cells on every successful transmission.
From a practical viewpoint and considering the typical loss probabilities, 
setting $n>1$ is not convenient because of the increased energy waste for idle listening.  
This aspect has been confirmed by an experimental evaluation of the \textit{2-open} strategy.

\subsection{A-open strategy}
The active-open (\textit{A-open}) strategy is an enhancement of \mbox{\textit{1-open}} that can be implemented only in devices 
capable to perform clear channel assessment (CCA), for instance by leveraging energy detection.
Similarly to \textit{1-open}, sleep commands in \textit{A-open} turn off reception on $\Nrx$ starting from cell $\cell{\overline{x+2}}$, 
while leaving it enabled in cell $\cell{x+1}$ (Fig~\ref{fig:comparison}.g). 
However, if the CCA mechanism on $\Nrx$ detects activity on air in cell $\cell{x+1}$, 
the next reserved cell ($\cell{x+2}$) remains enabled and the reception shut-off is postponed to $\cell{\overline{x+3}}$. 
This action is repeated until CCA detects a clear cell (e.g., a cell without any evidence of ongoing transmission).
Typically, the energy spent to perform CCA resembles what is wasted for idle listening.
In the worst case (when disturbance is repeatedly heard in subsequent scheduled cells), 
the same power consumption as conventional TSCH is experienced on $\Nrx$.

The example in Fig~\ref{fig:comparison}.i shows that, 
as in the \textit{1-open} approach in Fig~\ref{fig:comparison}.f, 
when $\frmsx{x}{t}=1$ the cell $x+1$ is left open on $\Nrx$
and can be used by $\Ntx$ for transmitting a frame.
In the \mbox{\textit{A-open}} case, however, 
if $\ack{x}=0$ and $\frmsx{x+1}{t-1}=0$
the detection of activity on air by CCA prevents PRIL from disabling reception in cell $x+2$, 
so that $\Nrx$ can hear the second retransmission.  
The most prominent advantage of this strategy is the total absence of useless talking, though the number of idle listening events increases as in the \textit{1-open} technique.

\subsection{PRIL operation}
\begin{figure}[]

\begin{center}
\footnotesize
\begin{alltt}
\textbf{Sender: N\(\sb{\textbf{TX}}\)}
 1: if NOT is_empty_buf() AND cell\(\sb{x}\)!=OFF:
 2:    F\(\sb{x}\)=dequeue_buf(); m=N\(\sb{tries}\)-1
 3:    t=get_next_t()
 4:    send(F\(\sb{x}\sp{.s=t}\))
 5:    while recv(A\(\sb{x}\))==0 AND m>0:
 6:        x++; m--; t--;
 7:        if t>0: send(F\(\sb{x}\sp{.s=t}\))
 8:        else: send(F\(\sb{x}\))
 9:    if A\(\sb{x}\)==1:
10:       cell\(\sb{x+1,...,x+t}\)=OFF
11:       if \underline{mode}!=\textit{closed}: cell\(\sb{x+1}\)=ON
         
\textbf{Receiver: N\(\sb{\textbf{RX}}\)}
12: if cell\(\sb{x}\)!=OFF:
13:    if recv(F\(\sb{x}\) OR F\(\sb{x}\sp{.s=t}\))==1:     # Any recv frame 
14:       send(A\(\sb{x}\))
15:       if F\(\sb{x}\sp{.s=t}\)==1:   # Only frame with sleep cmd
16:          cell\(\sb{x+1,...,x+t}\)=OFF
17:          if \underline{mode}!=\textit{closed}: cell\(\sb{x+1}\)=ON
18:    if \underline{mode}==\textit{A-open} AND CCA\(\sb{x}\)==1: cell\(\sb{x+1}\)=ON
\end{alltt}
\end{center}
  \caption{Pseudo-code for \textit{closed}, \textit{1-open}, and \textit{A-open} strategies (i.e., modes). The pseudo-code of $\Nrx$ is repeated for every single scheduled cell $\cell{x}$,
  while that of $\Ntx$ also encompasses all the subsequent scheduled cells possibly used for  retries upon transmission errors.}
  \label{fig:pseudo-code}
\end{figure}
In Fig.~\ref{fig:pseudo-code} the operation of three PRIL strategies is described using pseudo-code,
separately for $\Ntx$ and $\Nrx$. 
Variable \texttt{mode} can assume one of the values \textit{closed}, \textit{1-open}, and \mbox{\textit{A-open}}, 
and permits to differentiate the behavior of the strategies. 
For the sender $\Ntx$, in correspondence to the scheduled cell $\cell{x}$, 
when there is at least one pending frame and the cell was not switched off by PRIL (line 1)
a frame is extracted from the queue and the number of available tries $m$ is initialized. 
Function \texttt{get\_next\_t()} in line 3 predicts how many $\cell{}$ subsequent scheduled cells can be switched off because not needed for transmission, 
and returns the value $t$ to be included in the sleep command. 
Its actual implementation is out of the scope of this paper. 
After the initial transmission of frame $\frmsx{x}{t}$ in line 4, 
the frame is retransmitted until either the related $\ack{x}$ returns to $\Ntx$ or the number of allowed retries is over.
When $t$ reaches $0$ the sleep command is no longer included in the frame.
In sophisticate \texttt{get\_next\_t()} implementations, the $t$ value may change also between subsequent retries. 
In these cases, re-evaluation of $t$ has to be performed inside the \texttt{while} statement. 
If $\ack{x}$ notified the correct reception of the frame (line 9), 
the following $t$ cells related to $\cell{}$ are switched off. 
In lines 10 and 16, if $t<1$ the range $x+1,...,x+t$ identifies the empty set, and the command $\mathrm{cell}_{x+1,....x+t}=\mathrm{OFF}$ has no effect. Only for \textit{1-open} and \textit{A-open} the next cell $\cell{x+1}$ remains usable (line 11).

Regarding the receiver $\Nrx$, 
the pseudo-code reported in the figure is repeated for every scheduled cell $\cell{x}$ that was not switched off (line 12). 
After the correct reception of a frame, 
both those which include a sleep command and those which do not (line 13), 
the receiver sends an ACK frame back to $\Ntx$. 
If the received frame contains a \textit{sleep} command 
(test $\frmsx{x}{t}==1$ in line 15)
PRIL switches off the following $t$ cells (line 16). 
In the \textit{1-open} and \textit{A-open} case the next $\cell{x+1}$ cell is left usable.
Only in the \textit{A-open} case, 
if the channel is sensed busy (\texttt{CCA$_x$==1})
the next cell $\cell{x+1}$ is left open (line 18). 
Extending the algorithm to \textit{n-open} is simple: 
all it is needed is to change \texttt{cell$_{x+1}$=ON} with \texttt{cell$_{x+1,...,x+n}$=ON} (lines 11 and 17).

From the complexity viewpoint, strategies are comparable: 
in fact, their algorithmic implementations are quite similar. 
It is important to remark again that \textit{A-open} can be used only in devices provided with CCA support.

\section{Experimental Setup}
\label{sec:experimental_setup}
A discrete event simulator has been used to evaluate and compare the energy-saving techniques from the performance point of view. The simulator, called TSCH-predictor, is based on the \texttt{SimPy} framework. 
With respect to other tools like TSCH-Sim \cite{s20195663} and 6TiSCH \cite{6TiSCH_sim}, 
it benefits from a much simpler implementation that enables a quick and easy behavioral assessment of new algorithms proposed for TSCH. 
Regarding the features of TSCH-predictor that are of interest for this work, 
packet losses are modelled as Bernoulli trials with probabilities $\epsf$ and $\epsa$.
Different probabilities can be defined for uplinks and downlinks and, 
if needed, for every single channel. 
Statistics it provides about latency were cross-validated with measures obtained on real devices
in four experimental conditions. 
The discrepancy between real and simulated behavior, in terms of the average latency, 
was always below $\unit[6]{\%}$. 
Instead, statistics about power consumption were derived by counting the number of specific events 
(idle listening, frame transmission/reception) and by multiplying them by the 
related energy consumption derived from measurements obtained on real devices. 
More information about features and limitations of TSCH-predictor can be found in \cite{PRIL_etfa2021}. 

For the simulation campaigns performed in this paper,
the simulator was configured according to data collected on a real setup.
The values assigned to the main parameters are summarized in Table~\ref{tab:parameters}. 
The power consumption model derives from experiments carried out with Open-MoteSTM devices equipped with an Atmel AT86RF231 radiochip and an STM32F103RB 32-bit microcontroller \cite{2014-SJ-consumption}. 
In particular, the energy required by $\Ntx$ to transmit a $\unit[127]{B}$ data frame and receive the corresponding acknowledgement is about $\unit[485.7]{\mu J}$, 
while the energy needed by $\Nrx$ to receive such a frame and transmit the acknowledgement is about $\unit[651.0]{\mu J}$. 
Finally, the energy spent for idle listening is about $\unit[303.3]{\mu J}$.

\subsection{Estimation of ACK loss probability}
\label{subsec:ACK_prob}
Properly estimating the loss probability $\epsa$ for ACK frames plays 
an important role in this work. 
Hence, its evaluation relied on measurements performed on a real setup (the same used in \cite{9187609}). 
The testbed consists of OpenMote B devices running the OpenWSN operating system and the 6TiSCH protocol stack. 
A  Wi-Fi traffic pattern that mimics the variability of real traffic 
was purposely injected into the system to model environments suffering from non-negligible interference.
This simple expedient increased the frame loss probability and helped obtaining more realistic figures.
Generation of long traffic bursts was purposely avoided in interferers 
to make TSCH transmission attempts 
(for both data and ACK frames) as independent as possible.
The loss probabilities for ACK and data frames ($\epsa$ and $\epsf$)
were obtained from the \textit{Default} condition of \cite{9187609},
whose log is the longest one among those available in that report ($15$ days). 
In particular, the \texttt{ping} utility was used to repeatedly issue {ICMP echo} commands with period $\unit[120]{s}$:
\textit{ICMP echo request} packets were sent from the root node ($\Ntx$) to the leaf node ($\Nrx$), 
while \textit{ICMP echo replies} flowed in the opposite direction ($\cellr{}$). 

In the following we assume that the failure probability $\epsf$ is the same in the two directions $\cell{}$ and $\cellr{}$,
and the same holds for $\epsa$.
However, $\epsf$ and $\epsa$ may in general differ.
In the considered conditions the loss probability for data frames, computed using the method in \cite{9187609}, 
is $\epsf=\unit[12.6]{\%}$\footnote{Experimental data used to compute the loss probability values are included in the file \texttt{default-101-16-15days.dat}, which can be downloaded from \href{https://dx.doi.org/10.21227/fg62-bp39}{https://dx.doi.org/10.21227/fg62-bp39}.}.
Instead, the evaluation of $\epsa$ is not so straightforward. 
The probability $P_\mathrm{ACK}$ that, at the end of a frame transmission including up to ($\Ntries-1$) retries, the ACK frame is correctly received by $\Ntx$, under the assumption that spectrum conditions in the related cells are independent, is
\begin{align}
    \label{eq:Pnak}
    P_\mathrm{ACK} = 1-[\epsf + (1-\epsf)\epsa]^{\Ntries}.
\end{align}

Let us consider a single data frame exchange performed by means of confirmed transmission services.
The expected number $\overline N_\mathrm{txas}$ 
of ACK frames successfully received by $\Ntx$ is equal to $P_\mathrm{ACK}$.
In fact, whatever the sequence of attempts, the transmission procedure is terminated as soon as an ACK frame is correctly delivered to its destination. Moreover, when the retry limit is exceeded, no ACK frame is returned to $\Ntx$  and the overall probability of this event is $1-P_\mathrm{ACK}$.
Thus, 
$\overline N_\mathrm{txas} = 
1 \cdot P_\mathrm{ACK} + 0 \cdot (1-P_\mathrm{ACK}) = 
P_\mathrm{ACK}$.

The average number $\overline N_\mathrm{txaf}$ of ACK frames that are lost by $\Ntx$ can be computed as 
$\overline N_\mathrm{txaf} = \overline N_\mathrm{txa} - \overline N_\mathrm{txas}$, where $\overline N_\mathrm{txa}$ is the average number of ACK frames sent by $\Nrx$. 
$\overline N_\mathrm{txa}$ is equal to the average number $\overline N_\mathrm{rxf}$ of data frames received correctly by $\Nrx$.
It follows that the loss probability for ACK frames $\epsa$,
computed as the ratio $\overline  N_\mathrm{txaf}$ / $\overline N_\mathrm{txa}$,
is
\begin{align}
    \label{eq:epsaratio}
    \epsa 
    = \frac{\overline N_\mathrm{txaf}}{\overline N_\mathrm{txa}}
    = 1-\frac{\overline N_\mathrm{txas}}{\overline N_\mathrm{txa}}
    = 1-\frac{P_\mathrm{ACK}}{\overline N_\mathrm{rxf}}.
\end{align}

In the current OpenWSN implementation, 
IEEE 802.15.4 MAC layer sequence numbers are not exploited to discard duplicated frames. 
For this reason, every time an ACK frame is lost
the retransmitted frame is interpreted by the receiver as a new packet 
(in our case, an ICMP request or reply).
However, thanks to the sequence number field included in the ICMP message header,
the ICMP requestor (placed above the MAC layer in $\Ntx$) is able to detect duplicate ICMP replies.
They are labelled with the string ``\texttt{DUP!}'' in the \texttt{ping} log.

Every ICMP echo request sent by $\Ntx$ is received by $\Nrx$ $\overline N_\mathrm{rxf}$
times on average. 
Since deduplication is not performed by the MAC of the motes in the experimental testbed, 
this means that $\overline N_\mathrm{rxf}$ ICMP echo replies are sent back  to $\Ntx$, on average,
for every original ICMP echo request (accounting for possible losses and duplicates).
Replies, in their turn, may be lost or duplicated.
By assuming that transmissions concerning requests and replies are statistically independent, 
the average number of replies received by $\Ntx$ for every \texttt{ping} request it issues is equal to $\overline N_\mathrm{rxf}^2$.

The fraction of duplicate ICMP echo replies with respect to all the ICMP echo requests 
is $\alpha_\mathrm{dup}=\overline N_\mathrm{rxf}^2-1$.
This value can be estimated from the experimental logs collected for $\Ntx$ as
$\tilde \alpha_\mathrm{dup} = N_\mathrm{DUP} / N_\mathrm{ping}$,
where $N_\mathrm{DUP}$ is the number of log lines that are marked ``\texttt{DUP!}'',  
while $N_\mathrm{ping}$ is the overall number of issued ping requests.
Then  $\overline N_\mathrm{rxf}$ can be computed as
\begin{equation}
    \label{eq:Nrxf}
    \overline N_\mathrm{rxf} \simeq \sqrt{\tilde \alpha_\mathrm{dup} + 1}.
\end{equation}

$\epsa$ can be evaluated recursively by substituting \eqref{eq:Pnak} and \eqref{eq:Nrxf} in \eqref{eq:epsaratio}.
The starting value of $\epsa$ can be set equal to $\epsf$.
We have observed that less than $10$ interactions are enough to provide adequate accuracy. 
Using the values $\epsf=\unit[12.6]{\%}$, $N_\mathrm{DUP}=1967$, and $N_\mathrm{ping}=10800$, obtained from the experimental logs, 
the ACK loss probability, adopted in the remaining part of this work, is $\epsa=\unit[8.0255]{\%}$, 
which can be safely rounded to $\epsa=\unit[8.0]{\%}$.

\begin{table}[t]
  \caption{Main simulation parameters (default condition).}
  \label{tab:parameters}
  \footnotesize
  \begin{center}
  \includegraphics[width=1\columnwidth]{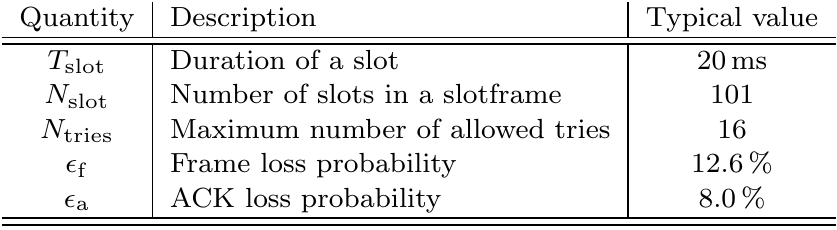}   
  \end{center}  
\end{table}

\section{Results}
\label{sec:results}

Reducing power consumption is the main goal of PRIL,
and the experimental campaign in Subsection~\ref{sec:power_consumption} compares the different strategies focusing on this aspect. 
Instead, the impact of the \textit{closed}, \textit{1-open}, \mbox{\textit{2-open}}, and \textit{A-open} strategies on latency 
is taken into account in Subsection~\ref{sec:latency}. 
Although responsiveness is not the primary goal of PRIL, the closer the latency is to conventional TSCH, the better the strategy.
This is particularly relevant when WSNs are deployed in industrial scenarios, where timings are often important.
Finally, Subsection~\ref{sec:taxonomy} provides a short recap of differences and similarities among strategies from several points of view.
To make comparison more valuable, we also included results about conventional TSCH 
(that is, without the adoption of any PRIL technique), which can be considered a sort of baseline.
Clearly, PRIL should be able to reduce power consumption of TSCH significantly in order to be attractive. 

\begin{table}
  \caption{Power consumption for different values of $\epsf$, $\epsa$, and $\Ntries{}$.}

  \label{tab:power}
  \footnotesize
  \begin{center}
  \includegraphics[width=1\columnwidth]{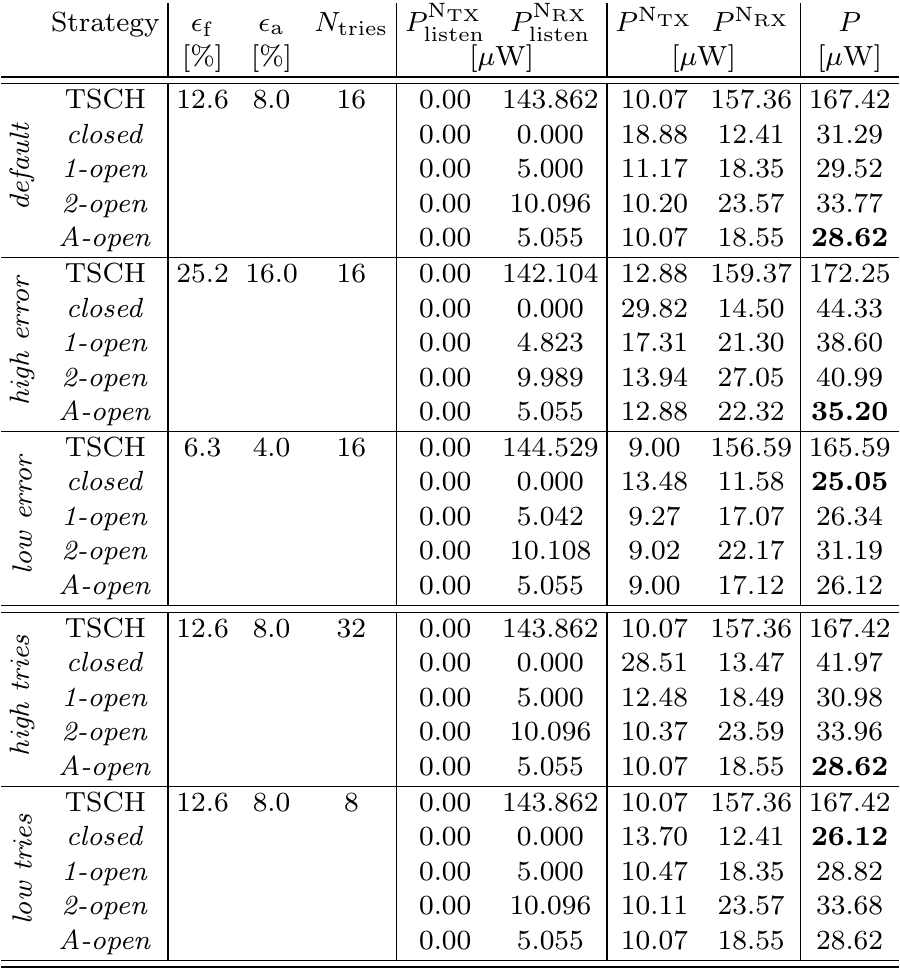}
  \end{center}
\end{table}

All experiments described in the following sections were carried out with two nodes ($\Ntx$ and $\Nrx$) and a 
cyclic packet flow with period $\Tapp$. 
The simulated duration was set to $\unit[10]{years}$ for all campaigns, 
and every experiment included at least $5256000$ packets (when $\Tapp=\unit[60]{s}$).

\begin{figure*}
\begin{center}
\includegraphics[width=2\columnwidth]{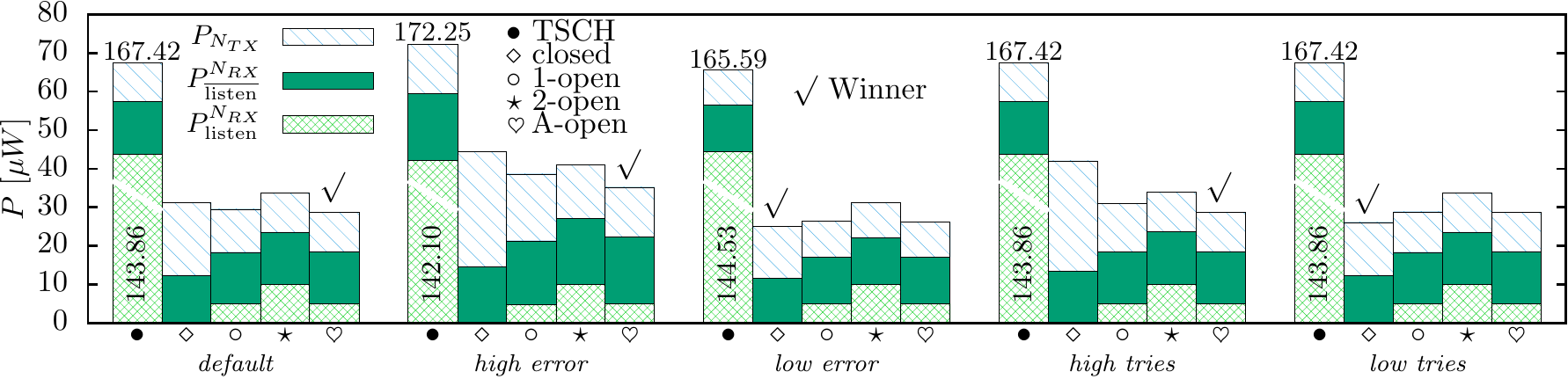}
\end{center}
\caption{Power consumption of PRIL strategies for different values of $\epsf$, $\epsa$, and $\Ntries{}$ 
(the lower the better, out of scale values are indicated explicitly).}
\label{fig:power}
\end{figure*}

\subsection{Power consumption}
\label{sec:power_consumption}
The $\Tapp=\unit[60]{s}$ period, corresponding to $\unit[3000]{slots}$, has been selected to prevent queuing phenomena in $\Ntx$. 
We have also carried out additional experiments aimed at analyzing the effect of queuing as shown in the following subsections.
Results obtained in experiments focusing on power consumption are summarized in Table~\ref{tab:power}.
In particular, $P$ is the total power consumption, 
$P^{\Ntx}$ and $P^{\Nrx}$ represent the power consumption on $\Ntx$ and $\Nrx$, respectively,
while
$P_{\mathrm{listen}}^{\Ntx}$ and $P_{\mathrm{listen}}^{\Nrx}$ refer to the specific contributions of idle listening.
For clarity, they are also plotted in Fig.~\ref{fig:power},
where 
$P$ is expressed as $P^{\Ntx}+P_{\overline{\mathrm{listen}}}^{\Nrx}+P_{\mathrm{listen}}^{\Nrx}$
and $P_{\overline{\mathrm{listen}}}^{\Nrx}$ concerns frame receptions.

The operating condition labelled \textit{default} in the table refers to the ``typical'' parameter settings we considered in this paper, where the values of $\epsf$, $\epsa$, and $\Ntries$ come directly from the real setup.
As expected, \textit{1-open} outperformed \textit{2-open} in all considered conditions. 
In fact, the penalty paid for leaving two cells ``open'' is higher than 
the benefit achieved by a better ability to prevent useless talking.
An interesting outcome is that \textit{A-open} always outperformed \textit{1-open}. 
This result is not surprising, as \textit{A-open} takes advantage from the ability of the wireless adapter to perform CCA. 
The other side of the coin is that its implementation is more complex than \textit{1-open} on some devices.

\begin{figure*}[t]
    \begin{center}
    \includegraphics[width=2\columnwidth]{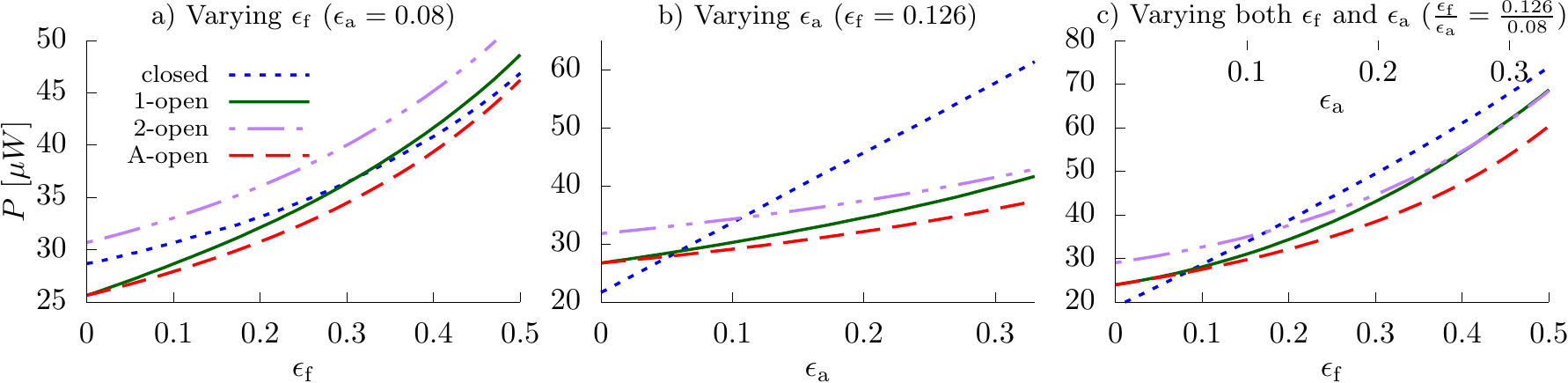}
    \end{center}
	\caption{Overall power consumption $P$ (considering both the transmitting and receiving sides) vs. failure rates $\epsf$ and $\epsa$ (for data and ACK frames).}
	\label{fig:power_probability}
\end{figure*} 
\begin{figure*}[t]
	\begin{center}
	\includegraphics[width=2\columnwidth]{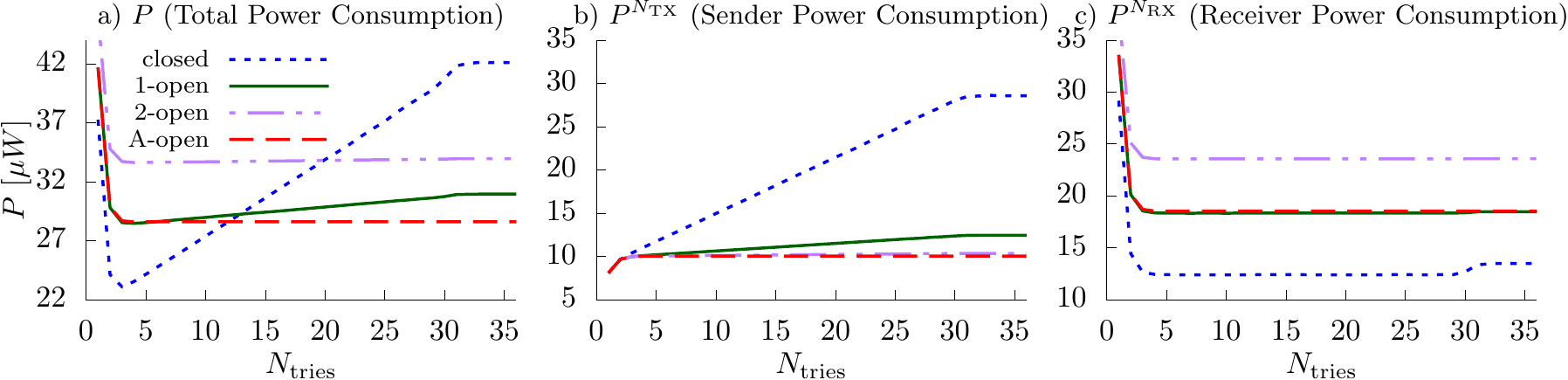}
	\end{center}
	\caption{Overall power consumption $P$ and contributions on the transmitter $P^{\Ntx}$ and the receiver $P^{\Nrx}$ vs. allowed transmission attempts $N_{\mathrm{tries}}$.}
	\label{fig:power_Ntries}
\end{figure*}

In the \textit{default} condition \textit{A-open} outperforms all other strategies, 
and the power saved with respect to \textit{1-open} and \textit{closed} is $\unit[0.90]{}$ and $\unit[2.67]{\mu W}$, 
respectively, corresponding to energy consumption reductions equal to $3.2\%$ and $9.3\%$. 
A significant difference between \textit{X-open} (\textit{n-open} and \textit{A-open}) and \textit{closed} strategies 
is that the former are optimal for the transmitter whereas the latter is the best for the receiver. 
This occurs because in the \textit{closed} strategy $\Ntx$ keeps retransmitting frames (on $\ack{x}=0$ events)
that cannot reach $\Nrx$ because the relevant cell is disabled, thus consuming a non-negligible amount of energy. 
On the contrary, in \textit{X-open} strategies a redundant number of cells is left open, which increases the energy wasted by $\Nrx$.

In the \textit{high error} condition reported in Table~\ref{tab:power} both $\epsf$ and $\epsa$ were doubled, whereas they were halved in the \textit{low error} case.
It is worth noting that the \textit{closed} strategy achieves best results in the \textit{low error} condition ($\epsf=6.3\%$ and $\epsa=4.0\%$): this is because the useless talking phenomenon is unlikely to occur when transmission attempts fail seldom.
In the two operating conditions at the bottom of Table~\ref{tab:power} $\Ntries$ was increased to $32$ (\textit{high tries}) and reduced to $8$ (\textit{low tries}), respectively.
For small $\Ntries$ values the \textit{closed} strategy is advantageous, because the amount of useless talking is limited by $\Ntries$. 
By contrast, for large values of $\Ntries$ the lowest energy consumption is achieved by the \textit{A-open} strategy.

When either the disturbance level (\textit{high error}) or the retry limit (\textit{high tries}) is increased, 
the gap between \textit{closed} and \textit{A-open}, in terms of power consumption, also grows. 
Thus, two additional sets of experiments were performed to investigate in more details 
the impact of the environment (disturbance) and network configuration (retry limit) on power consumption.
Results, summarized in Fig.~\ref{fig:power_probability}, concern the first set of experiments, 
where the total power consumption $P$ was studied by varying the failure probability.

In the leftmost plot (Fig. \ref{fig:power_probability}.a),
$\epsa$ is kept constant and equal to $8.0\%$ while $\epsf$  varies from $0$ to $50\%$. 
Curves obtained  for the four strategies show similar trends, 
and the consumption of \textit{closed} progressively approaches \textit{A-open}.

Differences in the curve shapes are evident in Fig.~\ref{fig:power_probability}.b, 
where $\epsf=12.6\%$ while $\epsa$  varies from $0$ to $33\%$. 
In this case, the total power consumption for the \textit{closed} strategy grows almost linearly with respect to $\epsa$
because of the larger amount of useless talking,
whereas it is less steep for \textit{X-open} strategies.

Finally, in Fig.~\ref{fig:power_probability}.c $\epsf$ and $\epsa$ are varied jointly according to a proportional law.
When the disturbance level is high \textit{A-open} implies lower energy consumption compared to the other strategies. 
In particular, \textit{A-open} is optimal when $\epsf>9\%$ and $\epsa>5.7\%$, 
otherwise \textit{closed} is the best choice. 

Fig.~\ref{fig:power_Ntries} shows the results obtained with the second set of experiments, 
where power consumption was analyzed by varying $\Ntries$ from $1$ to $36$ 
while failure probabilities were left fixed ($\epsf=12.6\%$ and $\epsa=8.0\%$).
The three plots show the overall power consumption $P$, as well as the contributions on the sender ($P^{\Ntx}$) and receiver 
 ($P^{\Nrx}$) side.
Once again, \textit{A-open} outperforms \textit{1-open} with respect to every metric ($P$, $P^{\Ntx}$, and $P^{\Nrx}$), while \textit{1-open} is better than \textit{2-open}.
Fig~\ref{fig:power_Ntries}.a shows that, as long as  $\Ntries \leq 12$, 
the \textit{closed} strategy is significantly less demanding than \textit{A-open}.
When $\Ntries \geq 3$  power consumption is almost independent from $\Ntries$ for both \textit{A-open} and \textit{2-open},
while for the other two strategies it increases linearly with different slopes (\textit{closed} is the steepest).

\begin{figure*}
\centering
    \includegraphics[width=2\columnwidth]{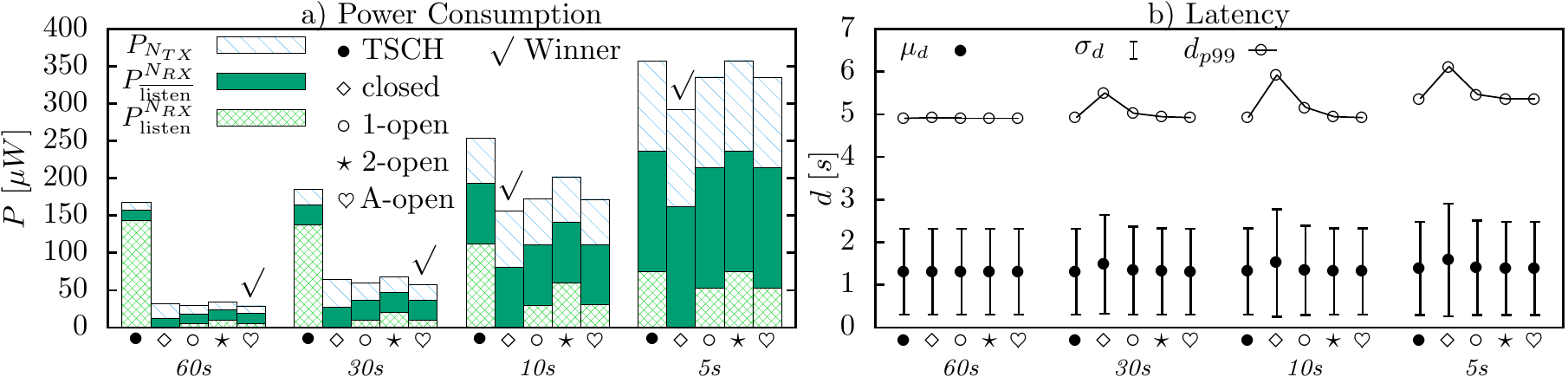}
\caption{Power consumption (dissected, on the left) 
and latency ($\mu_d \pm \sigma_d$ and $d_{p99}$, on the right) 
of PRIL strategies for different values of $\Tapp$.}
\label{fig:power_latency}
\end{figure*}

\begin{table*}[th]
  \caption{Comparison between the proposed PRIL strategies by varying $T_{\mathrm{app}}$ ($\epsf=12.6\%$, $\epsa=8\%$, $\Ntries=16$). 
  }
  
  \label{tab:latency}
  \small
  \begin{center}
  \includegraphics[width=2\columnwidth]{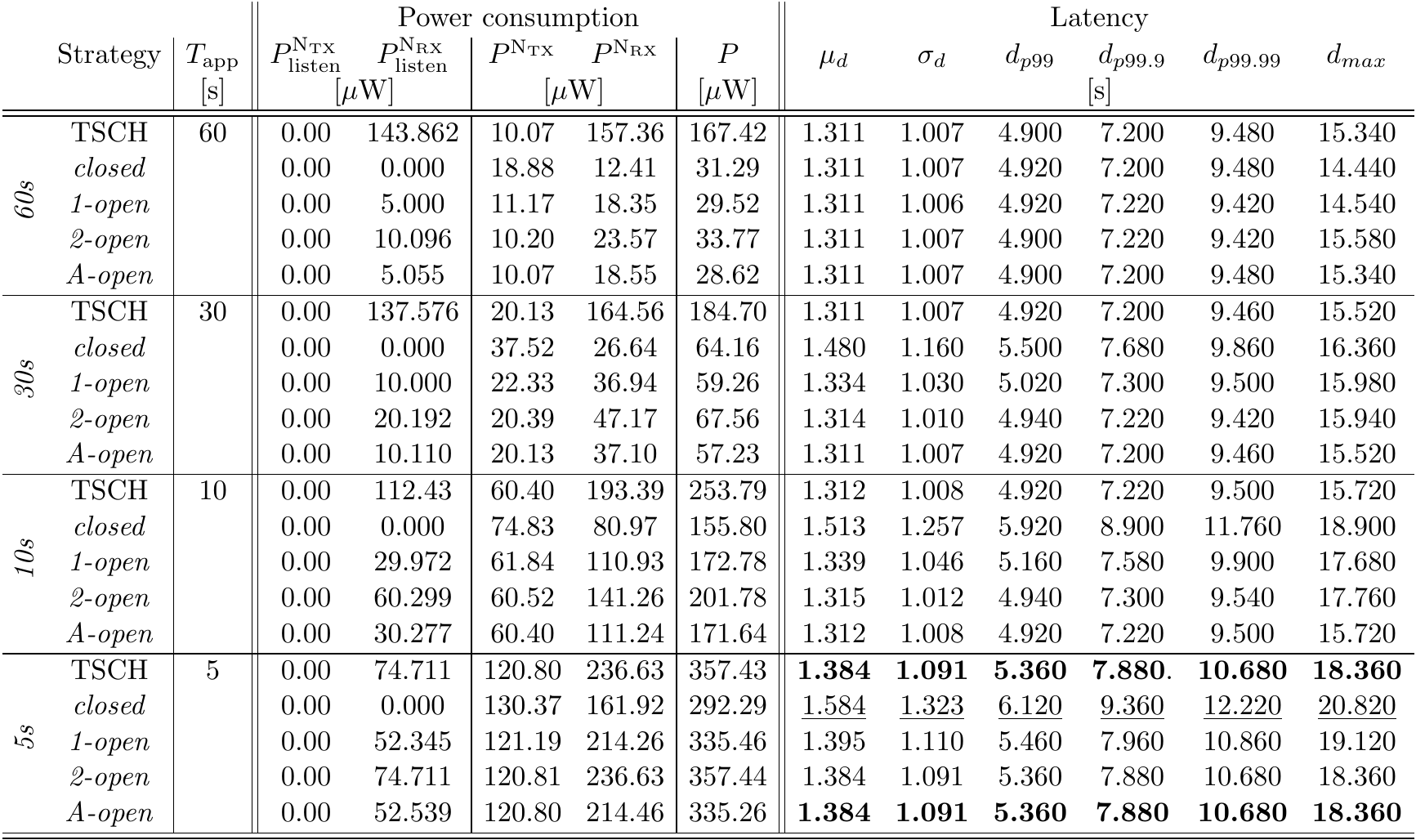}
  \end{center}
\end{table*}

It is worth observing that all strategies exhibit a peak when $\Ntries < 3$ 
because of the $P^{\Nrx}$ contribution (see Fig.~\ref{fig:power_Ntries}.c).
This behavior can be easily explained, 
since reducing the number of available retries increases the probability that all attempts for the same packet fail, 
thus preventing the sleep command from reaching $\Nrx$ and turning off the corresponding cells. 
Thus, the energy wasted for idle listening grows higher. 
Fig.~\ref{fig:power_Ntries}.c shows that the best strategy for $\Nrx$ is \textit{closed}, 
because it switches all the relevant cells off as soon as the sleep command is received. 
The other strategies leave one or more cells open after the reception of the command, and this contributes to energy waste.

Power consumption on $\Ntx$ is reported in Fig.~\ref{fig:power_Ntries}.b. 
With the exception of \textit{A-open}, it increases linearly with $\Ntries$ for all strategies
because of the useless talking phenomenon (when $\Ntx$ and $\Nrx$ have different views of the link state). 
Power consumption is independent from the number of available tries when $\Ntries \geq 30$.  
This is due to the value we selected for $\Tapp$ in the experiments, which is roughly equivalent to $30$ cells. 
In fact, the sleep command is configured to disable cells for a time interval equal to $\Tapp$, 
thus preventing useless talking from increasing further.

\subsection{Latency}
\label{sec:latency}

Latency is a key performance indicator for many industrial applications. 
For this reason a second experimental campaign was performed with the twofold goal of evaluating latency
as well as the effect of a reduction of $T_\mathrm{app}$ on both latency and power consumption. 
Four operating conditions have been considered, 
where $T_\mathrm{app}$ was set to $\unit[60]{s}$, $\unit[30]{s}$, $\unit[10]{s}$, and $\unit[5]{s}$, respectively, 
while the remaining parameters were the same as in the \textit{default} case described above.
Latency is defined as the time, as seen by the application, taken by a packet to travel from $\Ntx$ to $\Nrx$. 
In our case it includes the time spent by the packet in the queue of the sending node, 
the time to manage retransmissions, and a uniformly distributed waiting delay. 
The waiting delay is due to the lack of synchronization between packet generation at the application level 
and the slotframe boundaries defined by the network. 
In long experimental runs, variations of this delay are also affected 
by the clock skew between the synchronized network time and the local clock 
(derived from a free-running counter driven by the quartz oscillator of the node). 
Other delays, such as propagation times over the air, are negligible when compared to the above contributions.

Besides power consumption, results reported in Table~\ref{tab:latency} 
and in the plot on the right in Fig.~\ref{fig:power_latency} show the main statistical indices related to latency, 
namely mean ($\mu_d$), standard deviation ($\sigma_d$), 
99- ($d_{p99}$), 99.9- ($d_{p99.9}$) and 99.99-percentiles ($d_{p99.99}$),
and maximum ($d_{max}$).
Results labeled \textit{60s} in the upper rows of the table
highlight that latency is mostly the same for all the strategies analyzed in this paper.
When the generation period shrinks, as in the \textit{30s} case, 
latency for the \textit{closed} and \textit{n-open} strategies starts growing. 
In particular, the time between the generation of two subsequent packets is enough to prevent queuing phenomena in $\Ntx$ only when $\Tapp\geq 60s$.

In the other experimental conditions (\textit{10s} and \textit{5s}) latency increases for all strategies, 
including conventional TSCH. 
This is due to the larger average queuing time in $\Ntx$.
In the \textit{5s} case queuing effects are evident, 
and also TSCH and \mbox{\textit{A-open}} suffer from a latency increase. 
In general, \textit{closed} appears to be the slowest strategy, 
while \textit{A-open} is optimal because its latency is the same as conventional TSCH.
It is worth observing that, when the generation period is small enough, almost all scheduled cells are used to transmit/retransmit data frames, and the energy saved by PRIL is consequently lower.

\subsection{Optimal strategy}
\label{sec:taxonomy}

\begin{table}[t]
  \caption{Synopsis about strategies ($\blacksquare$ good, $\boxtimes$ acceptable, $\square$ subpar).}
  \label{tab:taxonomy}
  \small
  \begin{center}
  \includegraphics[width=1\columnwidth]{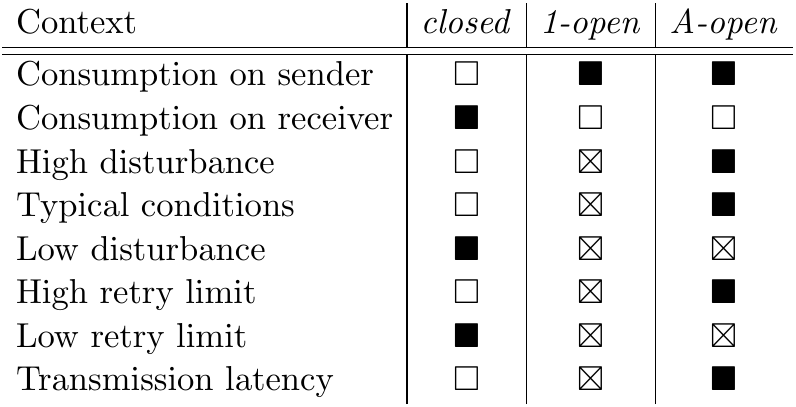}
  \end{center}
\end{table}

Simulations show that no strategy is optimal in every application context.  
For instance, \textit{A-open} outperforms other strategies in many cases, 
but in some circumstances the \textit{closed} behavior is better. 
Moreover, as expected, \textit{n-open} solutions do not provide any benefits in typical operating conditions when $n>1$.
Table~\ref{tab:taxonomy} recaps how
each strategy ranks in terms of power consumption in any given operating condition.
The first two lines refer to the two sides of the link in typical conditions. 
The last line concerns instead latency.  
A strategy was identified as ``good'', ``acceptable'', or ``subpar'',
depending on how it compares to the other strategies.

The table highlights that \mbox{\textit{A-open}} achieves the shortest latency and, most important, 
the lowest power consumption in critical operating conditions,
e.g., when the disturbance level is particularly high or the retry limit is increased to improve reliability. 
If CCA detects activity in an otherwise unused scheduled cell because of external interference, 
the \textit{A-open} strategy delays cell switching off to the next transmission opportunity, 
slightly increasing power consumption. 
This phenomenon was not modelled in the simulator because CCA is highly hardware-dependent (e.g., energy-based vs. carrier-based detection).
On the other hand, \textit{closed} ensures the best energy saving capability on the receiver side, 
and even for the whole link if the disturbance level and retry limit are low enough. 
Finally, \textit{1-open} can be profitably applied in the same conditions as \textit{A-open}, when the node is unable to perform CCA.

\section{Conclusions}
\label{sec:conclusions}

When dealing with proactive reduction of idle listening techniques in TSCH, the basic strategy, 
termed \textit{closed}, mitigates the problem and enables a consistent amount of energy to be conserved on the receiving side of a link.
However, it suffers from performance degradation when ACK frames get lost
and useless talking is experienced on the transmitting side.
The \mbox{\textit{1-open}} and \textit{A-open} strategies presented in this paper have been conceived to deal with this drawback, 
and have been checked against the \textit{closed} solution. 
Evaluation has been carried out by means of a discrete-event simulator developed ad hoc, 
and configured with parameters obtained from an experimental setup that includes real devices communicating over the air.

Results show that there is not a single winning strategy, 
as the behavior depends on specific operating conditions of the wireless network. 
Thus, we have identified the conditions under which every strategy offers the best performance, 
in terms of both power consumption  and latency.
This characterization shows that, in many realistic cases relevant for industrial contexts, 
\textit{A-open} outperforms the other strategies.

We deem that our analysis satisfactorily addresses and solves the problem caused by ACK losses in PRIL. 
Therefore, future work will focus on new and more effective implementations of PRIL techniques,
aimed at increasing the amount of saved energy by extending the currently available implementation (PRIL-F) 
to the multi-hop case.

\bibliographystyle{IEEEtran}
\bibliography{TII-21-5688}

\begin{IEEEbiography}[{\includegraphics[width=1in,height=1.25in,clip,keepaspectratio]{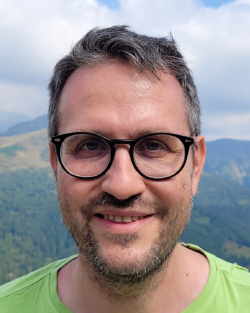}}]{Stefano Scanzio} (S’06-M’12-SM’22) received Laurea and Ph.D. degrees in computer science from Politecnico di Torino, Turin, Italy, in 2004 and 2008, respectively. From 2004 to 2009, he was with the Department of Computer Engineering, Politecnico di Torino, where he was involved in research on speech recognition and in classification methods and algorithms. Since 2009, he has been with the National Research Council of Italy, where he is currently a tenured researcher with the institute CNR-IEIIT. He teaches several courses on computer science at Politecnico di Torino. He has authored and co-authored more than 80 papers in international journals and conferences, in the areas of industrial communication systems, real-time networks, wireless networks, and artificial intelligence. He took part in the program and organizing committees of many international conferences of primary importance in his research areas. He received the 2017 Best Paper Award of the IEEE TRANSACTIONS ON INDUSTRIAL INFORMATICS and the Best Paper Awards for three papers presented at the IEEE Workshops on Factory Communication Systems in 2010, 2017, and 2019, and for a paper presented at the IEEE International Conference on Factory Communication Systems in 2020. He is an Associate Editor of the Ad Hoc Networks (Elsevier) and of the IEEE Access journals.
\end{IEEEbiography}

\begin{IEEEbiography}[{\includegraphics[width=1in,height=1.25in,clip,keepaspectratio]{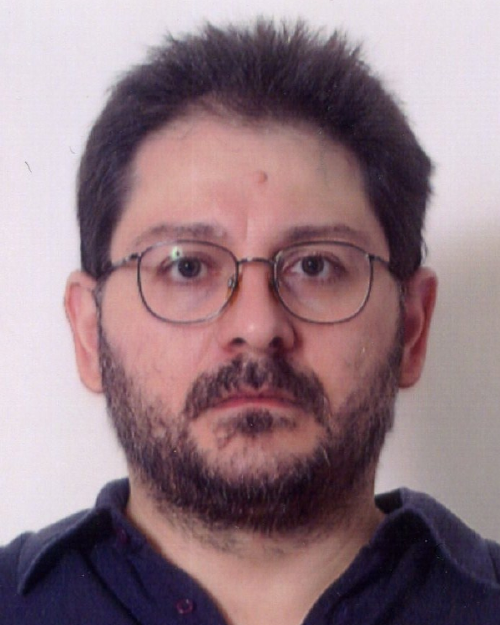}}]{Gianluca Cena}(SM’09) 
received the M.S. degree in electronic engineering and the Ph.D. degree in information and system engineering from the Politecnico di Torino, Italy, in 1991 and 1996, respectively.
Since 2005 he has been a Director of Research with the Institute of Electronics, Information Engineering and Telecommunications, National Research Council of Italy (CNR-IEIIT), Turin, Italy. 
His research interests include wired and wireless industrial communication systems, real-time protocols, and automotive networks. In these areas he has co-authored about 160 papers and one patent.
Dr. Cena was the recipient of the Best Paper Award of the IEEE TRANSACTIONS ON INDUSTRIAL INFORMATICS in 2017 and of the IEEE Workshop on Factory Communication Systems in 2004, 2010, 2017, 2019, and 2020.
He served as Program Co-Chairman of the IEEE Workshop on Factory Communication Systems in 2006 and 2008, and as Track Co-Chairman in six editions of the IEEE Conference on Emerging Technologies and Factory Automation. 
Since 2009 he has been serving as Associate Editor for the IEEE TRANSACTIONS ON INDUSTRIAL INFORMATICS.
\end{IEEEbiography}

\begin{IEEEbiography}[{\includegraphics[width=1in,height=1.25in,clip,keepaspectratio]{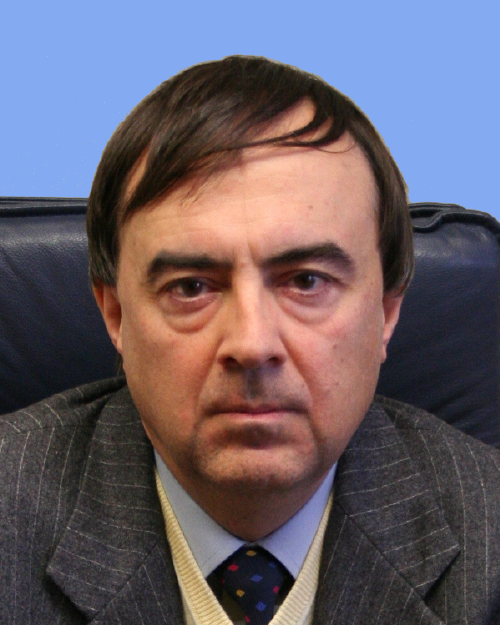}}]{Adriano Valenzano}(SM’09) received a Laurea degree magna cum laude in electronic engineering from Politecnico di Torino, Torino, Italy, in 1980. He is a director of research with the National Research Council of Italy (CNR) since 1991. He is currently with the Institute of Electronics, Computer and Telecommunication Engineering (IEIIT), Torino, Italy, where he is responsible for research concerning distributed computer systems, local area networks, and communication protocols. He has coauthored approximately 200 refereed journal and conference papers in the area of computer engineering. Dr. Valenzano is the recipient of the 2013 IEEE IES and ABB Lifetime Contribution to Factory Automation Award. He was also awarded for the best paper published in the IEEE TRANSACTIONS ON INDUSTRIAL INFORMATICS during 2016, and received the Best Paper Awards for the papers presented at the 5th, 8th, 13th, 15th and 16th IEEE Workshops on Factory Communication Systems (WFCS 2004, WFCS 2010, WFCS 2017, WFCS 2019 and WFCS 2020). Adriano Valenzano has served as a technical referee for several international journals and conferences, also taking part in the program committees of international events of primary importance. Since 2007, he has been serving as an Associate Editor for the IEEE TRANSACTIONS ON INDUSTRIAL INFORMATICS.
\end{IEEEbiography}

\cleardoublepage

\end{document}